\documentclass[useAMS,usenatbib]{mn2e}
\usepackage{amssymb}
\include{epsf}

\title[Microlensing and Lyman Alpha Forest]{Probing Sub-parsec Structure in the Lyman Alpha Forest with Gravitational Microlensing}
\author[B. J. Brewer and G. F. Lewis]{B. J. Brewer and G. F.
Lewis\thanks{E-mail: brewer@physics.usyd.edu.au (BJB), gfl@physics.usyd.edu.au (GFL)}\\
Institute of Astronomy, School of Physics, The University of Sydney, NSW 2006, Australia}
\begin{document}

\date{\today}

\pagerange{\pageref{firstpage}--\pageref{lastpage}} \pubyear{2004}

\maketitle

\label{firstpage}

\begin{abstract}
We present the results of microlens ray-tracing simulations showing the effect of absorbing material between a source quasar and a lensing galaxy in a gravitational lens system. We find that, in addition to brightness fluctuations due to microlensing, the strength of the absorption line relative to the continuum varies with time, with the properties of the variations depending on the structure of the absorbing material. We conclude that such variations will be measurable via UV spectroscopy of image A of the gravitationally lensed quasar Q2237+0305 if the Lyman Alpha clouds between the quasar and the lensing galaxy possess structure on scales smaller than $\sim 0.1$ pc. The time scale for the variations is on the order of order years to decades, although very short term variability can occur. While the Lyman alpha lines may not be accessible at all wavelengths, this approach is applicable to any absorption system, including metal lines.
\end{abstract}


\section{Introduction}The Lyman Alpha (Ly$_\alpha$) forest is an absorption phenomenon which is observed in the spectra of many high redshift quasars. It arises due to absorption by galactic and intergalactic neutral hydrogen between a quasar and an observer [see \citet{1998ARA&A..36..267R} for a recent review]. As the universe expands, photons from the quasar are redshifted, and therefore absorption by clouds at different cosmological distances cause absorption at different wavelengths in the spectrum. The resulting spectrum contains many absorption lines blueward of the Ly$_\alpha$ emission line of the quasar, and the structure of the absorbing material along the line of sight is imprinted on the spectrum \citep[e.g.][]{1996ApJ...472..509L}.

Numerical simulations of cosmological structure formation predict the large scale structure of the gas which can then be compared with observations of quasar spectra \citep{1999ApJ...511..521D}. The conclusions of these studies are that the Ly$_\alpha$ ``clouds" are mostly intergalactic structures, with the denser regions corresponding to the locations where galaxies form. The absorbing gas along the line of sight to a single quasar can be probed by comparing the relative positions and strengths of the absorption lines in the Ly$_\alpha$ forest in the spectrum of the quasar \citep{1980ApJS...42...41S}. The structure across the line of sight is harder to measure, although if two quasars that are nearby on the sky are observed, the fraction of Ly$_\alpha$ forest lines that are common to the two quasars can provide a measure of the extent of the clouds across the sky \citep{1982ApJ...256..374S, 2003MNRAS.341.1279R}.

Traditionally, the large scale structure of the Ly$_\alpha$ Forest has been the main focus of research, because of its cosmological significance \citep[e.g.][]{1999ApJ...518...24M}. Recently, the small scale structure (on kiloparsec scales) has also been investigated by several methods \citep{1999ApJ...515..500R, 2001ApJ...554..823R, 2001ApJ...562...76R, 2002ApJ...576...45R}; potentially revealing structure in and about the interstellar medium in young galaxies.

Gravitational lensing has previously been used to probe the size of absorbing clouds [note that if the mass associated with a Ly$_\alpha$ cloud is sufficiently large, it too may result in gravitational lensing effects, see \citet{2004astroph}]. When a background quasar (regarded as a point source) is lensed by a foreground galaxy, multiple images of the quasar may be obtained. The light rays responsible for the two different images have traversed slightly different regions of space in their journey to the observer. Hence, a measure of the size of the Ly$_\alpha$ absorbing clouds (down to kiloparsec scales) can be made by cross-correlating the the absorption lines in two images of the same quasar \citep{2004A&A...414...79E,2003AAS...20311301B}. This method is analogous to using a true pair of quasars, however the lensing supplies a ``pair'' of quasars that are close together on the sky (separated by $\sim$ arc seconds), allowing smaller scales to be investigated. \citet{1999ApJ...515..500R} used this method in the multiply imaged quasar Q1422+231 to trace MgII and CIV absorption clouds between the lens and the quasar down to scales of $\sim$ 26 pc, which is the smallest scale so far considered.

The very small scale ($\lesssim$ 1 parsec) structure of the Ly$_\alpha$ clouds is completely unknown (due to limits in observational techniques, and the fact that the resolution of Ly$_\alpha$ forest formation simulations is too coarse), and would be useful for deciding what the local equivalents of the Ly$_\alpha$ clouds are. Some (but not all) low redshift Ly$_\alpha$ clouds have been found to be associated with galaxies \citep{1995ApJ...441...51S}, suggesting that the structure of the Ly$_\alpha$clouds might mimic that of the interstellar medium. This is known to have structure on many scales, exhibiting fractal structure due to turbulence \citep{1997ApJ...477..196E}. In this paper we present simulations demonstrating how the very small, sub parsec scales of Ly$_\alpha$ clouds can be measured using a quasar that is subject to gravitational microlensing by a foreground galaxy.

\section{Background and Approach}

\begin{figure}
\begin{center}
\leavevmode
\epsfxsize=3in 
\epsfbox{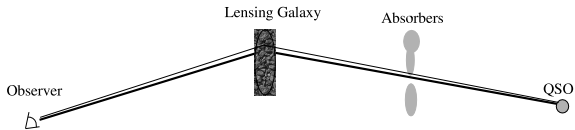}
\end{center}
\caption{As the relative positions of the observer, lens and quasar change (eg due to rotation and linear motion of the galaxy), the ray passes through a different region within the absorption clouds; causing the absorption line strength to change with time. This is similar to the changing positions of the micro-images with time \citep{1998ApJ...501..478L, 2004A&A...416...19T}.\label{diagram}}
\end{figure}

In gravitational microlensing, the lensing effects of individual stars in the lensing galaxy are considered [see \citet{1998LRR.....1...12W} for a review of gravitational lensing, including microlensing of quasars]. The cluster of `micro-images' that result are too small to be resolved \citep{1998ApJ...501..478L, 2004A&A...416...19T}, with the quasar appearing point-like, however brightness fluctuations can result as the relative positions of the source, observer and lensing stars change; this is the observational consequence of microlensing. Since the rays joining the observer to the quasar traverse different regions of space at different times, we anticipate that small scale structure in the absorption clouds will lead to time variations in the strength of the absorption line relative to the continuum (see Figure~\ref{diagram}).

Throughout this paper, we assume a cosmology with $\Omega_m=0.3$, $\Omega_\Lambda=0.7$ and $H_0 = 72$ km s$^{-1}$ Mpc$^{-1}$.

\subsection{Method}\label{method}
We made use of the microlensing code of \citet{1990PhDT.......180W}, modified to output the positions of the rays in the lens plane and the corresponding position in the source plane. The code uses the backward ray tracing method, where a rectangular grid of rays is fired into the lens plane, and the corresponding position in the source plane is calculated by summing the microlensing effects of many stars in the lensing galaxy. Using these positions, magnification maps (plots of the magnification of a point source as a function of position in the source plane) can be produced by binning the positions of the rays in the source plane (ie areas which collect a lot of rays have high magnification). For the simulations presented in this paper, typically $>10^9$ rays were traced between the observer and the source, ensuring that the resulting magnification maps, consisting of 1024x1024 binned pixels, did not suffer significant noise from Poisson sampling. A third plane, the ``cloud plane" was introduced, positioned between the lens plane and the source plane.

For any ray starting at a position $(x, y)$ in the lens plane and 
landing at a position $(x_s, y_s)$ in the source plane, the position 
$(x_p, y_p)$ at which that ray intersects the cloud plane is simply found by linear interpolation:

\begin{equation}
  \begin{array}{rcl}
    x_p & = & x + \lambda(x_s - x)\\
    y_p & = & y + \lambda(y_s - y)
  \end{array}
\label{interp}
\end{equation}

where $\lambda$ is a parameter which determines the location of the cloud plane. If $\lambda=0$, the cloud plane coincides with the lens plane, and can be used to model absorption clouds within the lensing galaxy \citep[e.g.][]{2003MNRAS.340..562L}. Alternatively, $\lambda=1$ corresponds to the case where the absorption clouds are located at the source. This situation has been used to investigate the effects of microlensing on broad absorption line quasars \citep{1998MNRAS.297...69L}. Note that $x, x_p,$ etc are scaled distances, but can be interpreted as the angular position of a point measured in units of the angular Einstein radius
\begin{equation}
\theta_0 = \sqrt{\frac{4GM_\odot}{c^2}\frac{D_{ls}}{D_{ol}D_{os}}}
\end{equation}
Hence, if a point in the cloud plane has scaled coordinates $(x_p,y_p)$, the physical coordinates (eg. measured in parsecs) are $(D_{op}\theta_0x_p,D_{op}\theta_0y_p)$, where $D_{op}$ is the angular diameter distance from the observer to the cloud plane. 

\begin{figure}
\begin{center}
\leavevmode
\epsfxsize=2in 
\epsfbox{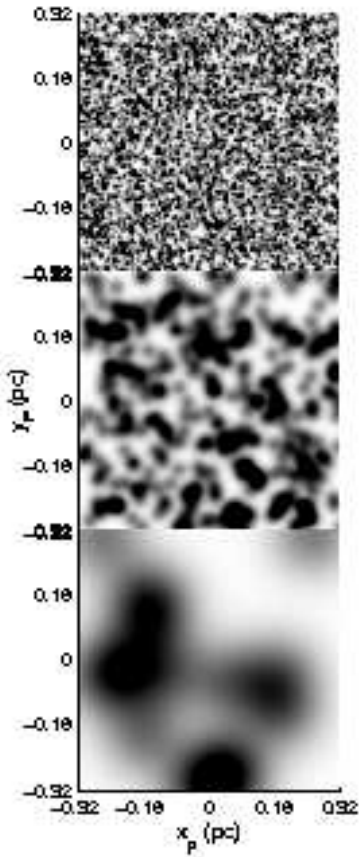}
\end{center}
\caption{The three different absorption cloud sizes which were investigated. The cloud sizes are 0.005 pc, 0.02 pc and 0.1 pc respectively. The plots show the absorption as a function of position in the cloud plane. Black corresponds to complete absorption ($A=1$), white corresponds to no absorption ($A=0$). Note that due to the external shear, the rays actually pass through a rectangular area in these images (see Figure~\ref{size}). These images of the clouds were for $\lambda$ = 0.2, the physical size of the plane was different in the other two cases to ensure that all the rays passed through the cloud plane.\label{clouds}}
\end{figure}

The clouds were described by an absorption function $A(x_p, y_p)$ defined over the cloud plane, which is the fraction of photons that are absorbed as a function of where the ray intersects the cloud plane. If a ray passes through a region where $A=0$, it is unaffected, but if a ray passes through a region where $A = 0.5$ say, then the ray is attenuated by 50\%. Gaussian clouds of the form $A=\frac{1}{2}e^{-r^2/b^2}$ were used, distributed randomly over the cloud plane. To investigate the effect of the cloud size $b$, we ran simulations with three different cloud sizes for each value of $\lambda$. The cloud sizes used were $b=0.005$ pc, $b=0.02$ pc and $b=0.1$ pc (for typical cloud configurations, see Figure~\ref{clouds}). The number of clouds in the cloud plane was chosen so that the average absorption per unit (physical) area of the cloud plane was approximately the same for all of the simulations (this was not exact due to a `clipping' effect, since the absorption can never be greater than 1 where several clouds coincide and their absorption values add). The mean value of the absorption was $\sim$ 0.5. To achieve computational efficiency, the absorption as a function of cloud plane position was stored as a 1024x1024 pixel image, and $A(x_p,y_p)$ was evaluated for each ray by looking up the appropriate pixel of the image. The physical size of the cloud plane was chosen to be large enough to ensure that all of the rays intersect it and can be affected by absorption.

\begin{figure}
\begin{center}
\leavevmode
\epsfxsize=3in 
\epsfbox{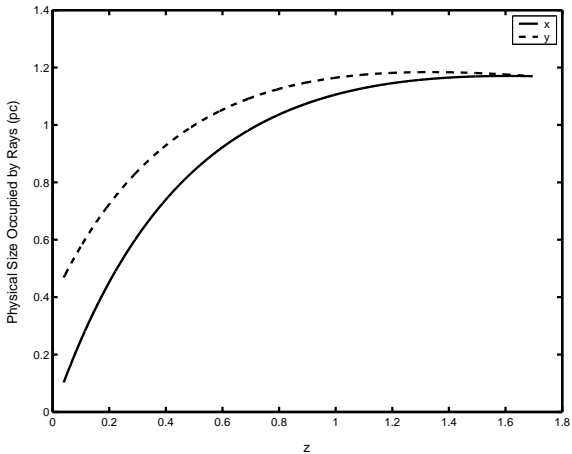}
\end{center}
\caption{Physical size occupied by the rays which land in the main region of the source plane, as a function of redshift (between the lens and the source, 0.0394 $<$ z $<$ 1.695). This relationship determines what area of the cloud plane is relevant for the rays which eventually land in the main square in the source plane. This explains why structure in the cloud plane is ``blown up'' in the line strength map more for clouds closer to the lens.\label{size}}
\end{figure}

For the purposes of this study, the microlensing parameters adopted were
those of image A of Q2237+0305, with a dimensionless surface mass density
(all in stars of mass 1 $\rm M_\odot$) of $\sigma=0.36$ and external shear
$\gamma=0.41$ \citep{1990ApJ...352..407W,1998MNRAS.295..488S}. The region
of the source plane that was considered was a square with side length 20
Einstein radii, ie a physical length 20$\theta_0D_{os}$, however this ray
bundle is only square at the source plane. It is approximately rectangular
for the rest of the journey, with side lengths as a function of redshift
shown in Figure~\ref{size}, which was calculated assuming that the lensing
galaxy was composed of smoothly distributed matter (see section~\ref{distances}). The redshift of the lens is $z = 0.0394$ and the redshift of the source is $z = 1.695$, which are the redshifts for Q2237+0305 and its lensing galaxy \citep{1985AJ.....90..691H}. For these redshifts, the physical size of an Einstein radius in the source plane is 5.85x10$^{-2}$ pc and the speed of the quasar across the source plane was assumed to be 600$\frac{D_{os}}{D_{ol}}$ km s$^{-1}$, or approximately one Einstein radius per decade \citep{1986A&A...166...36K,1998ApJ...501..478L}. We ignored the individual motions of stars in the lensing galaxy which, for a transverse velocity of 600 km s$^{-1}$ would cause a 10 per cent increase in the frequency of high magnification events \citep{1993ApJ...404..455K}. Ignoring this effect saves considerable computational effort.

The magnification maps provide a measure of the intensity of the source as a function of its position in the source plane. However, for a non-point source, the total observed intensity is the source profile multiplied by the magnification map, integrated over the source plane. Therefore the intensity as a function of source plane position for an extended source is given by the convolution of the magnification map with the source profile. The physical size of the continuum source in Q2237+0305 has been measured to be $\lesssim$ 7.8x10$^{-4}$ pc \citep{2002ApJ...579..127S}, which is $\sim 0.25$ pixels wide in the magnification maps. Therefore, we ignored finite source effects, taking a single pixel to represent a point-like source.

\begin{figure}
\begin{center}
\leavevmode
\epsfxsize=3in 
\epsfbox{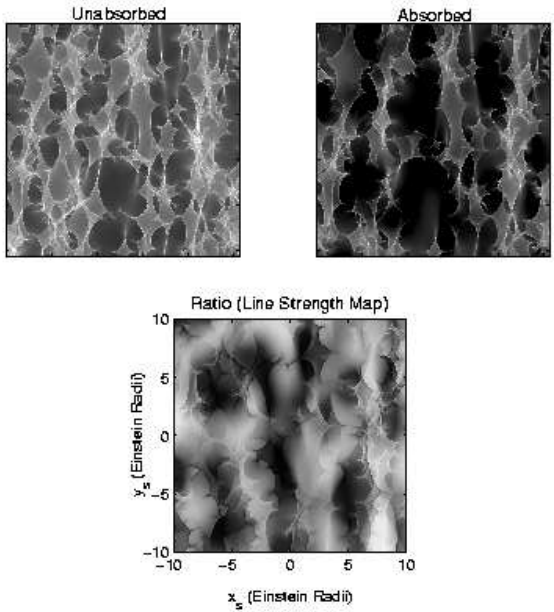}
\end{center}
\caption{Examples of magnification maps showing the magnification as a function of position of the source (the cloud plane position was $\lambda=0.5$ and the cloud size was 0.02 pc for these plots). The  first is the unabsorbed magnification map for the continuum, the second is the absorbed magnification map for the flux in the Ly$_\alpha$ absorption line. The third is the ratio of the two, and measures the strength of the absorption line as a function of quasar position (light areas denote a weak absorption line, dark areas indicate a deep absorption line).\label{examples}}
\end{figure}

For each simulation, a pair of magnification maps was produced. One was the usual magnification map calculated using the final locations of the rays, neglecting absorption, representing the magnification of the unabsorbed (continuum) wavelengths of the quasar spectrum. An absorbed magnification map was also produced, using the attenuated values of the rays. This magnification map applies to the absorbed part of the spectrum, so measures the flux at the bottom of the absorption line as a function of source position. A third map was produced, a ``line strength map'', which measures the flux at the bottom of the continuum relative to the flux of the continuum as a function of source position. It is obtained by dividing the absorbed magnification map by the unabsorbed map (see Figure~\ref{examples}).

\begin{figure}
\begin{center}
\leavevmode
\epsfxsize=3in 
\epsfbox{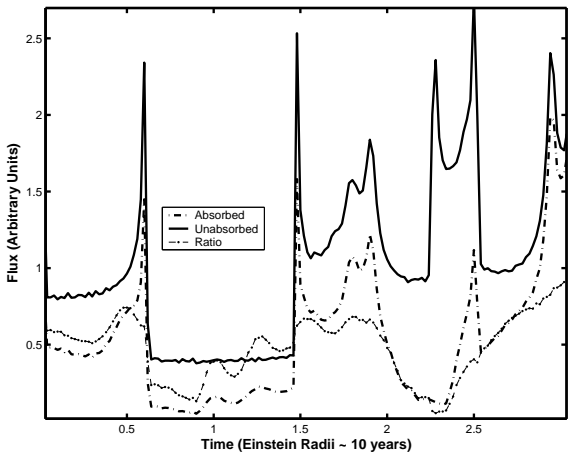}
\end{center}
\caption{A typical light curve showing how the flux in the absorption line (dashed line) and the continuum (solid line) vary with time (produced by taking a horizontal track across the magnification maps). The structure of the absorption clouds produces variations in the line flux relative to the continuum (dotted line). The time units depend on the velocity of the quasar, for a typical velocity of a few hundred km s$^{-1}$ this plot spans a time scale of several decades for Q2237+0305. The y-axis was normalised to a mean of 1, corresponding to the theoretical mean magnification, $\mu_{th} = \frac{1}{(1-\sigma)^2 - \gamma^2}.$\label{lightcurve}}
\end{figure}

A sample light curve showing the flux in the continuum and the absorption line, as well as the strength of the absorption line as functions of time is shown in Figure~\ref{lightcurve}. This light curve was taken from the same magnification maps shown in Figure~\ref{examples} by taking a one dimensional horizontal cut across the middle of the magnification map. Note that some (but not all) of the sudden changes in continuum brightness (due to the source crossing a caustic in the source plane) are accompanied by sudden changes in the line strength. Also, for times when the continuum magnification is approximately constant (eg from times $\sim 0.6 - 1.4$ in Figure~\ref{lightcurve}), the line strength time series may still vary, reflecting the structure in the absorption clouds.

\subsection{Distances and Cosmology}\label{distances}
\begin{figure}
\begin{center}
\leavevmode
\epsfxsize=3in 
\epsfbox{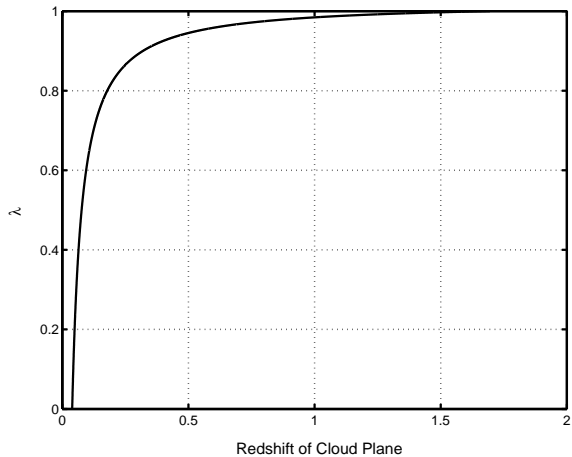}
\end{center}
\caption{The interpolation parameter $\lambda$ as a function of the redshift of the cloud plane for Q2237+0305 ($z_l = 0.0394, z_s = 1.695$). \label{z_dist}}
\end{figure}

The scaled lens mapping equations (which relate a position in the lens plane, or the image, to a position in the source plane) are
\begin{equation}
  \begin{array}{rcl}
    x_s & = & x - \alpha_x(x,y)\\
    y_s & = & y - \alpha_y(x,y)
  \end{array}
\end{equation}
where $\alpha_x$ and $\alpha_y$ are the scaled deflection angles. Substituting these equations into Equation~\ref{interp}, converting back to unscaled coordinates and using the unscaled lens equation \citep{1990PhDT.......180W}, it can be shown that $\lambda$ is given by
\begin{equation}
	\lambda = \frac{D_{os}D_{lp}}{D_{op}D_{ls}}
\end{equation}
where $D_{os}$, $D_{lp}$, $D_{op}$ and $D_{ls}$ are the angular diameter distances from the observer to source plane, lens plane to cloud plane, observer to cloud plane and lens plane to source plane respectively. The redshifts were converted to angular diameter distances using the assumed cosmology. The resulting plot of $\lambda$ vs z is shown in figure~\ref{z_dist}. Note that the flatness of this relationship over a wide redshift range implies that the results for $\lambda = 0.95$ are approximately applicable for clouds over a very broad redshift range. Hence, the very small scales of the Ly$_\alpha$ forest could be probed over a wide redshift range ($z \sim 0.5$ to $z \sim 1.6$) by comparing observations with the high $\lambda$ simulations. Also, the physical size $X_p$ of any feature in the cloud plane with scaled size $x_p$ is simply
\begin{equation}
X_p = D_{op}\theta_0x_p
\end{equation}
In this paper, we investigate the cases $\lambda = 0.2, \lambda = 0.5$ and $\lambda = 0.95$. For these values of $\lambda$, the corresponding redshifts for the clouds are $z = 0.049, 0.077$ and 0.53 (although the latter result would also work well for clouds at higher redshifts). Absorption by neutral hydrogen at these redshifts would produce an absorption line in the Q2237+0305 spectrum at wavelengths of 127.5 nm, 130.9 nm and 186.0 nm respectively. While the light from the quasar may suffer significant absorption below the Lyman limit [corresponding to z $\lesssim$1 ($<$ 245.0 nm)] which will influence the study of the Ly$_\alpha$ lines, the analysis presented in this paper is equally applicable to the ubiquitous metal lines that are also seen in quasar spectra and can arise at lower redshifts \citep{2002ApJ...576...45R}.

\section{Results}\label{results}
\subsection{Line Strength Maps}

\begin{figure}
\begin{center}
\leavevmode
\epsfxsize=3in 
\epsfbox{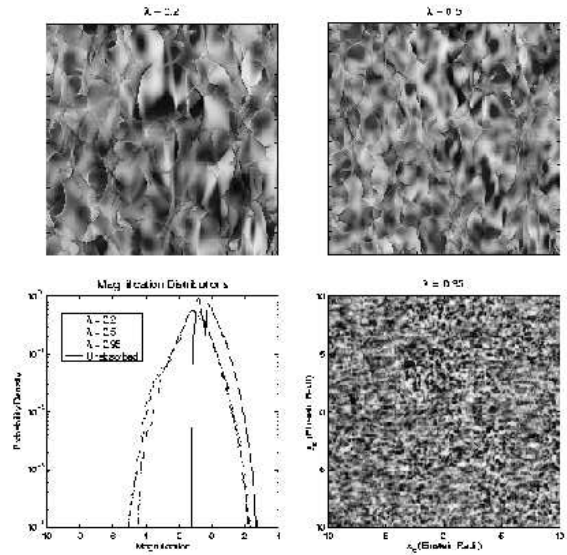}
\end{center}
\caption{Line strength maps and magnification PDFs for the ``small'' cloud size 0.005 pc, for clouds at three different redshifts. Changes occur on scales much smaller than an Einstein Radius, with the fastest variations occurring for absorption clouds close to the lens. The units for the magnification PDFs are magnitudes, with 0 being the mean continuum magnification.\label{panel1}}
\end{figure}

\begin{figure}
\begin{center}
\leavevmode
\epsfxsize=3in 
\epsfbox{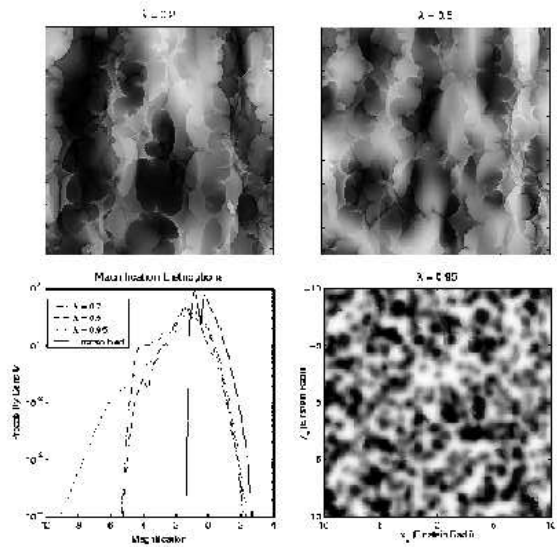}
\end{center}
\caption{As for Figure~\ref{panel1}, but for the ``medium" cloud size 0.02 pc.\label{panel2}}
\end{figure}

\begin{figure}
\begin{center}
\leavevmode
\epsfxsize=3in 
\epsfbox{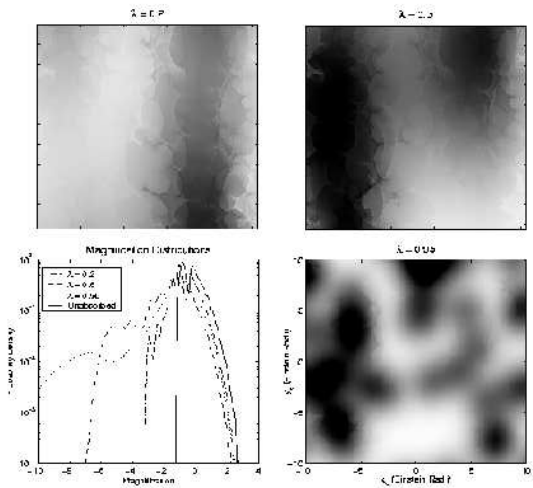}
\end{center}
\caption{Line strength maps and magnification PDFs for the ``large'' cloud size 0.1 pc.\label{panel3}}
\end{figure}

If the absorbing material was uniform over the small scales probed by microlensing, then the flux in the absorption line and the flux in the continuum would rise and fall together as the source moved across the source plane. In this case, the line strength map would be uniform, and a time series of measured values of the absorption line strength relative to the continuum would also be flat. Also, the distribution of magnifications [probability density function (PDF) for the magnification of a source placed at a random position in the source plane, see \citet{1992ApJ...386...19W, 1995MNRAS.276..103L}] for the flux in the line would be the same as that for the continuum, but dimmed by a constant absorption value. Figures~\ref{panel1}, ~\ref{panel2} and ~\ref{panel3} show line strength maps for all of the cases investigated in this paper, together with magnification distributions for the continuum and the flux in the absorption line. The magnification distributions for the absorbed part of the spectrum are all essentially the same, with the most likely value having the flux in the absorption line about 1 magnitude fainter than the flux in the continuum. The discrepancies in the faint end for the larger cloud sizes are likely to be due to sampling error, with only a small number of clouds affecting the final magnification map.

For the $\lambda = 0.2$ and $\lambda = 0.5$ cases, the structure in the line strength map is complex. The caustic network from the microlensing is evident, so rapid brightness changes due to the source crossing a caustic will usually be accompanied by sudden changes in line strength, albeit not as drastic as the sudden changes in the continuum brightness. Since the rays which landed in the relevant area of the source plane occupy a smaller region of space for lower $\lambda$, only a small region of the cloud plane gets imprinted on the line strength map. Therefore, for a particular cloud size, faster variations in the line strength occur for clouds close to the source. The effect of the external shear in stretching both the magnification map and the cloud related structures in the line strength map in the vertical direction is clearly visible.

When the absorption clouds are at $\lambda = 0.95$ (close to the source), the imprint of the caustic structure becomes much less obvious in the line strength map. The line strength map in this case approximately resembles the cloud structure, so if $\lambda$ is sufficiently high, the absorption can be approximated as being at the source plane, and line strength variations directly probe the structure absorbing material. As Figure~\ref{z_dist} demonstrates, this approximation is applicable over a wide range of redshifts.

For the low values of $\lambda$, the rays which land in the region of interest of the source plane have only crossed a small area of the cloud plane, so the features in the line strength map are quite large. In contrast, the $\lambda=0.95$ line strength map shows that fast variations would occur in this case.

For the large 0.1 pc clouds (Figure~\ref{panel3}), the line strength maps contain features that are typically several Einstein radii across, corresponding to time scales of several decades for any changes to be observed. Hence, this cloud size is approaching an upper limit of the scales that could be probed by this method. The caustic structure is still present, but the same structure is also present in the line strength maps for the other cloud sizes, so this cannot be used to distinguish different cloud sizes.

\begin{figure*}
\begin{center}
\leavevmode
\epsfxsize=5in 
\epsfbox{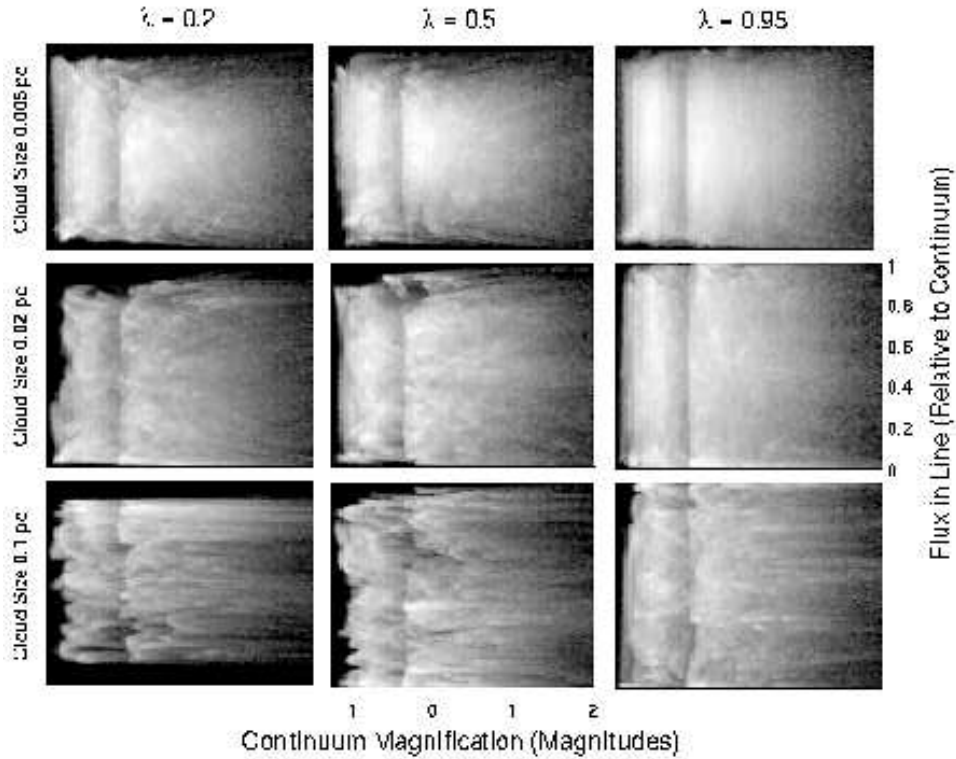}
\end{center}
\caption{Two dimensional histograms of the line flux relative to the continuum (y axis) versus the continuum magnification (x axis) for each of the cases presented in this paper. The greyscale indicates the log of the relative probability of observing a particular continuum magnification and line strength; a strong correlation would appear as a narrow white ridge line. There are no strong correlations between the line strength and the continuum magnification, despite the same patterns being apparent in the maps. The dip in the magnification distribution for the continuum \citep{1995MNRAS.276..103L, 1992ApJ...386...30R} is evident in these plots.\label{corr}}
\end{figure*}

The fact that the same caustic network is present in the magnification map for the continuum and the line strength map suggests that a correlation may be present. For example, areas where the line strength would be strong may be associated with areas where the quasar has a high magnification, or vice versa. To investigate this, histograms were produced, plotting the line strength versus continuum magnification (one point for each pixel in the maps). These are shown in Figure~\ref{corr}. An observation of a single light curve would sample regions from one of these distributions, possibly making different cloud sizes distinguishable. In each case the magnification probability distribution for the unobscured quasar source is the same. Clearly, Figure~\ref{corr} reveals that there is no strong correlation between the continuum magnification and the line strength relative to the continuum. Even though the imprint of the caustic network on the line strength maps also implies that rapid changes in the quasar brightness will be accompanied by significant changes in the line strength, there is no tendency for the changes to be correlated. Therefore, while spectral monitoring of the quasar during a high magnification event may make line strength variations measurable, they do not provide much information on the nature of the absorption clouds. Longer term spectral monitoring would be necessary to achieve this.

\subsection{Time Scales}

The usual way to quantify the time scale for microlensing variations is by computing the autocorrelation functions of a sample of light curves \citep{1994A&A...288....1S,1994A&A...288...19S}. In this case, it is the line strength variations which are being investigated, so the line strength map was used instead of the magnification map. We used a slightly different approach to measure the typical time scale for line strength variations as a function of $\lambda$ and the size of the absorbing clouds. For a given line strength map, a sample of pairs of points a fixed distance apart was selected at random, and the mean difference $<L_1 - L_2>$ in the line strength values between the two points was calculated. 

For example, if the pairs of points are close together relative to the size of the variations, the line strength is similar for those two points, so the mean difference is small. In this way, a plot of mean difference versus separation was produced. The separation at which the mean difference is half of its maximum value was taken to be the measure of the length scale of features in the line strength map.

\begin{table}
 \centering
 \begin{minipage}{80mm}
  \caption{Time scales for line strength variation as a function of cloud size and cloud position. The units are Einstein Radii, which take approximately 9.5 years to cross.\label{timescales}}
  \begin{tabular}{@{}llrrrrlrlr@{}}
  \hline

    & $\lambda$ = 0.2 & $\lambda$ = 0.5 & $\lambda$ = 0.95 \\
 \hline
 Small & 0.33 & 0.24 & 0.09 \\
 Medium & 1.14 & 0.81 & 0.37 \\
 Large & 3.9 & 6.4 & 1.7 \\
\hline
\end{tabular}
\end{minipage}
\end{table}

Using the above method, the time scales for the variations for each case presented in this paper were calculated (see Table~\ref{timescales}). The units in the table are Einstein Radii; for the assumed transverse velocity of 600 km s$^{-1}$ for the lensing galaxy, an Einstein Radius is crossed in about 9.5 years (see Section~\ref{method}). 

An examination of Table~\ref{timescales} reveals two clear trends. Firstly, as expected, variability in the line strength is most rapid for the smallest clouds. Additionally, it is apparent that the time scale of variability for a fixed physical cloud size decreases with increasing $\lambda$, and hence redshift. This implies that for a single microlensed quasar possessing identical structure throughout the Ly$_\alpha$ forest, the line variability will be most rapid for those absorption lines resulting from clouds closest to the quasar (remembering the non-linear relationship between $\lambda$ and z presented in Figure~\ref{z_dist}).

The fastest variations occur for small clouds and clouds close to the source, with the variations being observable over a small number of decades for both the small and medium cloud sizes at any position between the lens and the source. The anomalous value for the largest clouds is probably due to a sampling effect, as the source plane was not large enough to be affected by many clouds.

\begin{figure}
\begin{center}
\leavevmode
\epsfxsize=3in 
\epsfbox{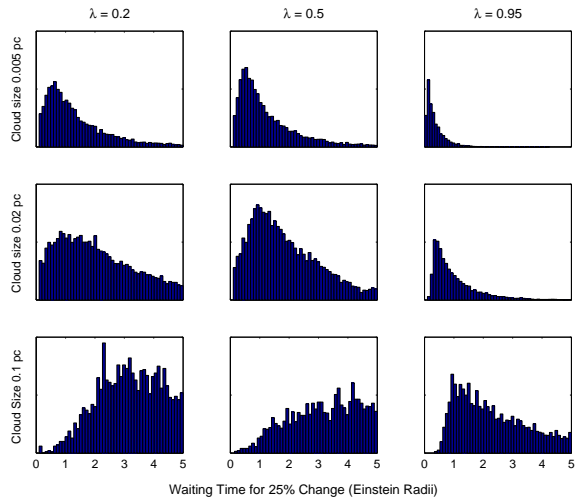}
\end{center}
\caption{Distribution of the waiting time for a 25 per cent change in the line strength. The x-axes are all on the same scale, with the time units being Einstein Radii or about 9.5 years. The medians of these distributions are shown in Table~\ref{medians}.\label{waittime}}
\end{figure}

For this effect to be measurable, there must be a significant change in the line strength over a time scale which is not too long. An absolute change of 0.25 in the line strength ought to be detectable even in low quality spectra. Recall that our measure of line strength is the ratio of the flux at the bottom of the absorption line to the flux in the nearby continuum. We measured the distribution of the waiting times for a change of 0.25 in the line strength. This was done by selecting a random starting point in the line strength map, and heading off in a random direction, moving until the line strength has changed by 0.25. The distribution of waiting times for this to occur is shown in Figure~\ref{waittime}, with the corresponding medians in Table~\ref{medians}. As expected, the waiting time tends to be shortest for small clouds close to the source, and the effect is observable in a reasonable time scale for all but the largest clouds.

\begin{table}
 \centering
 \begin{minipage}{80mm}
  \caption{Medians of the waiting time distributions in Figure~\ref{waittime}, in units of Einstein Radii (about 9.5 years). Note that the results for the large clouds are underestimated because the high tails of the distributions are were neglected. The peak of the distributions are evident though.\label{medians}}
  \begin{tabular}{@{}llrrrrlrlr@{}}
  \hline

    & $\lambda$ = 0.2 & $\lambda$ = 0.5 & $\lambda$ = 0.95 \\
 \hline
 Small & 0.86 & 0.76 & 0.26 \\
 Medium & 1.80 & 1.60 & 0.70 \\
 Large & 3.07 & 3.27 & 2.07 \\
\hline
\end{tabular}
\end{minipage}
\end{table}

In an unlensed quasar, the transverse velocities of the quasar and the Ly$_{\alpha}$ clouds would also produce variations in the strength of the absorption lines. For a transverse velocity of 600 km s$^{-1}$, the time taken to traverse 1 pc of the cloud plane (with clouds at redshifts $z = 0.049, 0.077$ and 0.53) are 1700, 1750 and 2500 years respectively. Clearly, these are much longer than the corresponding time scales for the lensed system. For example, even for the small clouds of size $\sim$ 0.005 pc, the time scale is $\gtrsim$ 8.5 years, approximately a factor of 10 slower than the microlensed case. Thus, microlensing provides the opportunity to study the small scale structure on a reasonable time scale. For the large clouds, the effect is even more drastic. The crossing time of a cloud is $\sim$ 2000 years, roughly 3 orders of magnitude longer than in the microlensed case.

\section{Conclusion}
The cosmological distribution of neutral hydrogen gas leaves an imprint on the spectrum of background quasars, in the form of a series of absorption lines. This phenomenon is thought to be related to the large scale flow of gas and formation of galaxies and clusters. While strong constraints have been placed on the larger scale properties of the Ly$_\alpha$ clouds, virtually nothing is known on smaller scales, where gas may possess similar turbulent properties to the local interstellar medium, or be collapsing to form stars. In this paper, we have examined the influence of gravitational microlensing on the paths taken by light rays impinging on a distribution of Ly$_\alpha$ clouds lying between the lens and the source. For a single macro-image, the population of micro-images samples differing absorption regions, with the net absorption seen (in a single line) being the average of that experienced by each ray. As the stars move, the configuration of micro-images changes, sampling different regions of the absorption clouds and causing temporal variability of the absorption line strength. 

In this study we considered clouds of three different sizes and at three different locations between the lens and the source. Quantitatively, it was shown that the fastest variations in the absorption line strength occur for the smallest clouds, and for clouds close to the source. We also found that, due to the geometry of the universe, the variations in the absorption lines in the spectrum of Q2237+0305 due to clouds at redshifts $\gtrsim 0.5$ would have approximately the same effect as identical clouds which are just in front of the source. At all redshifts, the caustic structure evident in the microlensing magnification map is also present in the absorption line strength map. These discontinuities can result in very rapid changes in the absorption line strength, with a corresponding change in the continuum brightness. However, a more detailed examination of the of the relationship between the continuum magnification and the line strength shows no apparent correlation between the two.

An analysis of the variability timescales reveals that significant changes in line strength due to the effects of gravitational microlensing would be visible on time scales ranging from years to decades, dependent on the size on the size of the absorption clouds and their location behind the lens; this is at least an order of magnitude faster than clouds simply drifting across the line of sight to a distant quasar. Hence, in order to measure this effect, long term spectroscopic monitoring of the individual images of microlensed quasars is required. Such observations have already been proposed for studies of quasar structure \citep{2002ApJ...576..640A, 2002ApJ...579..127S, 2004MNRAS.348...24L}.

This study has only considered Ly$_\alpha$ clouds in image A of Q2237+0305, which is a positive parity image. Other images of Q2237+0305 have negative parity, corresponding to one of the curves in Figure~\ref{size} passing through zero. Furthermore, this study has assumed that the absorbing clouds possess no kinematic substructure, and hence the resultant variability applies to the total strength of the line; for this a simple equivalent width measure in relatively low resulution spectra would suffice. Nevertheless, any real distribution of gas will exhibit some kinematic substructure in the form of rotation, infall, or simple clumping. This would result in not only variability in the total strength of an absorption line, but also in its profile [see \citep{1998MNRAS.297...69L}]. To detect line profile variability, higher resolution spectra would be required. We will discuss these issues in future contributions.

\section*{Acknowledgments}
This work was undertaken as part of an honours project at The University of Sydney. BJB was supported by a scholarship from the School of Physics and Cadbury-Schweppes, and would like to thank all the regulars of the SSSF ({\tt http://www2b.abc.net.au/science/k2/stn/}) for not telling me off when I was supposed to be doing work. GFL would like to thank The Rolling Stones for pointing out that you can't always get what you want.

\label{lastpage}

\begin{thebibliography}{99}

\bibitem[\protect\citeauthoryear{Abajas et al.}{2002}]{2002ApJ...576..640A} 
Abajas C., Mediavilla E., Mu{\~ n}oz J.~A., Popovi{\' c} L.~{\v C}., Oscoz 
A., 2002, ApJ, 576, 640 

\bibitem[\protect\citeauthoryear{Becker, Sargent, \& 
Rauch}{2003}]{2003AAS...20311301B} Becker G., Sargent W.~L.~W., Rauch M., 
2003, AAS, 203,  

\bibitem[\protect\citeauthoryear{Dav{\' e} et 
al.}{1999}]{1999ApJ...511..521D} Dav{\' e} R., Hernquist L., Katz N., 
Weinberg D.~H., 1999, ApJ, 511, 521 

\bibitem[\protect\citeauthoryear{Ellison et 
al.}{2004}]{2004A&A...414...79E} Ellison S.~L., Ibata R., Pettini M., Lewis 
G.~F., Aracil B., Petitjean P., Srianand R., 2004, A\&A, 414, 79 

\bibitem[\protect\citeauthoryear{Ellison et 
al.}{1999}]{1999PASP..111..946E} Ellison S.~L., Lewis G.~F., Pettini M., 
Sargent W.~L.~W., Chaffee F.~H., Foltz C.~B., Rauch M., Irwin M.~J., 1999, 
PASP, 111, 946 

\bibitem[\protect\citeauthoryear{Elmegreen}{1997}]{1997ApJ...477..196E} 
Elmegreen B.~G., 1997, ApJ, 477, 196 

\bibitem[\protect\citeauthoryear{Huchra et al.}{1985}]{1985AJ.....90..691H} 
Huchra J., Gorenstein M., Kent S., Shapiro I., Smith G., Horine E., Perley 
R., 1985, AJ, 90, 691

\bibitem[\protect\citeauthoryear{Kayser, Refsdal, \& 
Stabell}{1986}]{1986A&A...166...36K} Kayser R., Refsdal S., Stabell R., 
1986, A\&A, 166, 36 

\bibitem[\protect\citeauthoryear{Kundic \& 
Wambsganss}{1993}]{1993ApJ...404..455K} Kundic T., Wambsganss J., 1993, 
ApJ, 404, 455 

\bibitem[\protect\citeauthoryear{Lewis \& 
Belle}{1998}]{1998MNRAS.297...69L} Lewis G.~F., Belle K.~E., 1998, MNRAS, 
297, 69 

\bibitem[\protect\citeauthoryear{Lewis \& 
Ibata}{2004}]{2004MNRAS.348...24L} Lewis G.~F., Ibata R.~A., 2004, MNRAS, 
348, 24 

\bibitem[\protect\citeauthoryear{Lewis \& 
Ibata}{2003}]{2003MNRAS.340..562L} Lewis G.~F., Ibata R.~A., 2003, MNRAS, 
340, 562 

\bibitem[\protect\citeauthoryear{Lewis \& 
Ibata}{1998}]{1998ApJ...501..478L} Lewis G.~F., Ibata R.~A., 1998, ApJ, 
501, 478 

\bibitem[\protect\citeauthoryear{Lewis \& 
Irwin}{1995}]{1995MNRAS.276..103L} Lewis G.~F., Irwin M.~J., 1995, MNRAS, 
276, 103 

\bibitem[\protect\citeauthoryear{Lu et al.}{1996}]{1996ApJ...472..509L} Lu 
L., Sargent W.~L.~W., Womble D.~S., Takada-Hidai M., 1996, ApJ, 472, 509

\bibitem[\protect\citeauthoryear{M\'{e}nard}{2004}]{2004astroph}
M\'{e}nard, B. 2004, {\it astro-ph/0408276}, submitted to ApJ

\bibitem[\protect\citeauthoryear{McDonald \& Miralda-Escud{\' 
e}}{1999}]{1999ApJ...518...24M} McDonald P., Miralda-Escud{\' e} J., 1999, 
ApJ, 518, 24 

\bibitem[\protect\citeauthoryear{Rauch et al.}{1992}]{1992ApJ...386...30R} 
Rauch K.~P., Mao S., Wambsganss J., Paczynski B., 1992, ApJ, 386, 30 

\bibitem[\protect\citeauthoryear{Rauch}{1998}]{1998ARA&A..36..267R} Rauch 
M., 1998, ARA\&A, 36, 267 

\bibitem[\protect\citeauthoryear{Rauch et al.}{2002}]{2002ApJ...576...45R} 
Rauch M., Sargent W.~L.~W., Barlow T.~A., Simcoe R.~A., 2002, ApJ, 576, 45 

\bibitem[\protect\citeauthoryear{Rauch et al.}{2001}]{2001ApJ...562...76R} 
Rauch M., Sargent W.~L.~W., Barlow T.~A., Carswell R.~F., 2001, ApJ, 562, 
76 

\bibitem[\protect\citeauthoryear{Rauch, Sargent, \& 
Barlow}{2001}]{2001ApJ...554..823R} Rauch M., Sargent W.~L.~W., Barlow 
T.~A., 2001, ApJ, 554, 823

\bibitem[\protect\citeauthoryear{Rauch, Sargent, \& 
Barlow}{1999}]{1999ApJ...515..500R} Rauch M., Sargent W.~L.~W., Barlow 
T.~A., 1999, ApJ, 515, 500 

\bibitem[\protect\citeauthoryear{Rollinde et 
al.}{2003}]{2003MNRAS.341.1279R} Rollinde E., Petitjean P., Pichon C., 
Colombi S., Aracil B., D'Odorico V., Haehnelt M.~G., 2003, MNRAS, 341, 1279 

\bibitem[\protect\citeauthoryear{Salpeter \& 
Hoffman}{1995}]{1995ApJ...441...51S} Salpeter E.~E., Hoffman G.~L., 1995, 
ApJ, 441, 51 

\bibitem[\protect\citeauthoryear{Sargent et 
al.}{1980}]{1980ApJS...42...41S} Sargent W.~L.~W., Young P.~J., Boksenberg 
A., Tytler D., 1980, ApJS, 42, 41 

\bibitem[\protect\citeauthoryear{Sargent, Young, \& 
Schneider}{1982}]{1982ApJ...256..374S} Sargent W.~L.~W., Young P., 
Schneider D.~P., 1982, ApJ, 256, 374 

\bibitem[\protect\citeauthoryear{Schmidt, Webster, \& 
Lewis}{1998}]{1998MNRAS.295..488S} Schmidt R., Webster R.~L., Lewis G.~F., 
1998, MNRAS, 295, 488 

\bibitem[\protect\citeauthoryear{Seitz, Wambsganss, \& 
Schneider}{1994}]{1994A&A...288...19S} Seitz C., Wambsganss J., Schneider 
P., 1994, A\&A, 288, 19 

\bibitem[\protect\citeauthoryear{Seitz \& 
Schneider}{1994}]{1994A&A...288....1S} Seitz C., Schneider P., 1994, A\&A, 
288, 1 

\bibitem[\protect\citeauthoryear{Shalyapin et 
al.}{2002}]{2002ApJ...579..127S} Shalyapin V.~N., Goicoechea L.~J., Alcalde 
D., Mediavilla E., Mu{\~ n}oz J.~A., Gil-Merino R., 2002, ApJ, 579, 127 

\bibitem[\protect\citeauthoryear{Treyer \& 
Wambsganss}{2004}]{2004A&A...416...19T} Treyer M., Wambsganss J., 2004, 
A\&A, 416, 19 

\bibitem[\protect\citeauthoryear{Wambsganss}{1990}]{1990PhDT.......180W} 
Wambsganss J., 1990, PhD Thesis, Max-Plank-Institut f\"{u}r Physik und Astrophysik, M\"{u}nchen

\bibitem[\protect\citeauthoryear{Wambsganss}{1992}]{1992ApJ...386...19W} 
Wambsganss J., 1992, ApJ, 386, 19 

\bibitem[\protect\citeauthoryear{Wambsganss}{1998}]{1998LRR.....1...12W} 
Wambsganss J., 1998, Living Reviews in Relativity, 1, 12 

\bibitem[\protect\citeauthoryear{Wambsganss, Paczynski, \& 
Katz}{1990}]{1990ApJ...352..407W} Wambsganss J., Paczynski B., Katz N., 
1990, ApJ, 352, 407

\end{thebibliography}
\end{document}